\documentstyle[12pt,aaspp,epsf]{article}
\def\v4{$V_4$}
\def\i4{$I_4$}
\def\asec{\rlap{$^{\prime\prime}$}.\hbox to 2pt{}}
\def\amin{\rlap{$^\prime$}.\hbox to 1pt{}}
\def\mag{\hskip 0.5pt\rlap{$^{\rm m}$}{\hskip 1.5pt.}\hbox to 2.0pt{}}

\def\etal{et al.\ \rm}

\def\square{\vbox to 7pt{\boxit{\hbox to 0pt{}}}\hskip 1pt}

\def\footnoterule{\kern-3pt\hrule\kern 2.6pt}

\newbox\grsign \setbox\grsign=\hbox{$>$} \newdimen\grdimen \grdimen=\ht\grsign
\newbox\simlessbox \newbox\simgreatbox
\setbox\simgreatbox=\hbox{\raise.5ex\hbox{$>$}\llap
     {\lower.5ex\hbox{$\sim$}}}\ht1=\grdimen\dp1=0pt
\setbox\simlessbox=\hbox{\raise.5ex\hbox{$<$}\llap
     {\lower.5ex\hbox{$\sim$}}}\ht2=\grdimen\dp2=0pt
\def\simgreat{\mathrel{\copy\simgreatbox}}
\def\simless{\mathrel{\copy\simlessbox}}

\def\vol#1  {{{#1}{\rm,}\ }}
\def\araa{{ARA\&A}, }
\def\aj{{AJ}, }  %Astronomical Journal%
\def\apj{{ApJ}, } %Astrophysical Journal%
\def\apjs{{ApJS}, } %Astrophysical Journal Supplements%
\def\apjl{{ApJ}, } %Astrophysical Journal Letters%
\def\pasp{{PASP}, }  %Publications of the Astronomical%
\def\mnras{{MNRAS}, } %Monthly Notices of the Royal%
\def\aa{{A\&A}, }     %Astronomy & Astrophysics%
\def\aasup{{A\&AS}, } %A & A Supplements%
\def\hi{\noindent \hangindent=2.5em}

\def\clock{\count0=\time \divide\count0 by 60
     \count1=\count0 \multiply\count1 by -60 \advance\count1 by \time
     \number\count0:\ifnum\count1<10{0\number\count1}\else\number\count1\fi}

\begin{document}
%\draft

\title{ The Palomar Distant Cluster Survey :  \\
        II. The Cluster Profiles}

\author{Lori M. Lubin\altaffilmark{1}}
\affil{Princeton University Observatory, Peyton Hall, Princeton NJ 08544}

\author{Marc Postman\altaffilmark{2}}
\affil{Space Telescope Science Institute\altaffilmark{3}, 3700 San 
Martin Dr., Baltimore, MD 21218}

\altaffiltext{1}{Present Address : Observatories of the Carnegie Institution
of Washington, 813 Santa Barbara St., Pasadena, CA 91101}
\altaffiltext{2}{Guest Observer, Palomar Observatory}
\altaffiltext{3}{Space Telescope Science Institute is operated by the
Association of Universities for Research in Astronomy, Inc.,
under contract to the National Aeronautics and Space Administration.}

\vskip 4 cm
\centerline{Accepted for publication in the {\it Astronomical Journal}}
%\centerline{15 August 1995; resubmitted 18 December 1995}

\vfill
\eject

% \baselineskip 22pt
%
% ABSTRACT
%

\begin{abstract}
We present a study of the surface density profiles of the clusters of
galaxies from the Palomar Distant Cluster Survey (Postman et al.
1996).  The survey contains a total of 79 clusters of galaxies,
covering the estimated redshift range of $0.2 \simless z \simless
1.2$.  We have analyzed the richest clusters in this sample and find
that the typical Palomar cluster has a surface density profile of
$r^{-1.4}$ ($r \ge 0.10~h^{-1}~{\rm Mpc}$) and a core radius of
$0.05~h^{-1}~{\rm Mpc}$. There may be an indication that the slope of
the surface density profile steepens with increasing redshift, though
the observational uncertainty is at present too large to be
conclusive. Our cluster population is inconsistent at a 99.9\%
confidence level with a population of azimuthally symmetric clusters.
\end{abstract}

{\it Subject headings}: galaxies: clustering; cosmology: observations

\section{Introduction}

Clusters of galaxies provide a powerful probe of the nature of galaxy
formation and the origin of large scale structure in the universe.
Having a well-studied sample of intermediate and high-redshift
clusters is vital to understanding the evolution of galaxies and large
scale structure. Because the diversity of cluster properties and the
effects of evolution are significant at high redshift (Gunn \&
Dressler 1988; Bower et al. 1994; Castander et al. 1994; Postman et
al. 1996), it is essential to sample many distant clusters with
different properties. Large scale simulations have shown that cluster
properties such as profile shape and substructure can constrain the
nature of large scale structure formation theories and the mass
density of the universe (Evrard et al. 1993; Crone, Evrard \&
Richstone 1995; Jing et al. 1995; Tyson \& Fischer 1995).  Detailed
studies of these observational parameters have already been made in
some individual, intermediate redshift clusters.  Using weak lensing
to map the mass distribution in several clusters, Smail et al. (1995)
have studied two X-ray luminous clusters 1455+22 ($z = 0.26$) and
0016+16 ($z = 0.55$), Tyson \& Fischer (1995) have studied Abell 1689
($z = 0.18$), and Squires \etal (1995) have studied Abell 2218 ($z =
0.175$).  All four clusters show moderate to extreme degrees of
structure.  However, the distribution of structure in mass, light, and
X-rays are all well correlated at radii larger than $100~h^{-1}~{\rm
kpc}$. The projected mass density (and light) profile of A1689 can be
well approximated by a power-law of $r^{-1.4 \pm 0.2}$, while the
projected total mass, gas, and light surface densities of A2218 are
consistent with an isothermal sphere ($r^{-1}$), though perhaps
slightly steeper. Smail et al. (1995) found that, while the galaxies
are very good tracers of the mass, they are less concentrated, the
respective core radii for 1455+22 being $r_{c}^{mass} =
50^{+40}_{-25}~h^{-1}~{\rm kpc}$ and $r_{c}^{gal} =
90^{+35}_{-25}~h^{-1}~{\rm kpc}$.  A multi-color photometric study of
three clusters, A3284 ($z = 0.15$), A3305 ($z = 0.16$), and A1942 ($z
= 0.23$), also yield consistent core radii of $r_{c}^{gal} \approx
120,~100,~{\rm and}~120~h^{-1}~{\rm kpc}$, respectively (Molinari et
al.\ 1994).

In this paper, we expand on these detailed analyses of individual
intermediate redshift clusters by studying a large sample of
intermediate and high redshift clusters from the Palomar Distant
Cluster Survey (hereafter PDCS; Postman et al. 1996).  We examine both
the individual and global properties of the cluster profiles,
specifically the cluster profile slope, core radius, and degree of
asymmetry.  Because the PDCS uses a completely objective and automated
algorithm to detect clusters, it provides us with the largest,
statistically complete sample of distant clusters.  In addition, the
selection biases due to the detection technique can be well quantified
through simple Monte-Carlo simulations.

In \S 2 of this paper we briefly describe the PDCS and the cluster
sample used in this analysis.  The parameters of the individual
cluster profiles are discussed in \S 3. Composite cluster profiles as
a function of redshift are presented in \S 4. Cluster morphology is
discussed in \S 5.  We present a preliminary comparison of our results
to large-scale cosmological simulations in \S 6 and summarize the
results in \S 7.

\section{The Cluster Sample}

The cluster sample is derived from an optical/near IR survey with the
4-shooter CCD camera on the Palomar 5 meter telescope.  The cluster
catalog is the subject of the first paper in this series (Postman et
al. 1996; hereafter Paper I).  However, we discuss briefly the aspects
of this catalog which are necessary for the following analysis.

The Palomar Distant Cluster Survey was conducted in two broad band
filters, the F555W and F785LP of HST's Wide Field/Planetary Camera. We
denote these bands \v4 (F555W) and \i4 (F785LP) according to the
convention in Paper I.  The area covered consists of five one--square
degree areas around the sky, all at galactic latitudes of 30 degrees
or greater.  The data are complete to \v4 $= 23.8$ and \i4 $= 22.5$.
A matched filter algorithm was used to objectively identify the
cluster candidates by using positional and photometric data
simultaneously.  This technique is likely to be more robust than
previous optical selections which simply looked for surface density
enhancements, a method which can be significantly affected by
superposition effects ({\it e.g.,} Gunn, Hoessel \& Oke 1986; Couch et
al.\ 1991). An advantage of this technique is that redshift estimates
of the cluster candidates are produced as a byproduct of the matched
filter; the main disadvantage is that we must assume a particular form
for the cluster luminosity function (for the flux filter) and cluster
radial profile (for the radial filter). The radial filter $P(r)$ and
the flux filter $L(m)$ are given by

\begin{eqnarray}
P(r) = & {1 \over {\sqrt{1 + (r/r_{c})^{2}}}} - {1 \over {\sqrt{1 + (r_{co}/r_c)^{2}}}} & \mbox{if $r < r_{co}$} \nonumber \\
        & 0 & \mbox{              otherwise}
\end{eqnarray}

\begin{equation}
L(m) = {{\phi(m-m^{*})~10^{-0.4(m-m^{*})}}\over{b(m)}}
\end{equation}
 
\noindent $P(r)$ is an azimuthally symmetric cluster surface density
profile which has a characteristic core radius ($r_{c}$) and which
falls off at large radii as $r^{-1}$.  The function is explicitly cut
off at an arbitrary cutoff radius ($r_{co}$). We have chosen $r_{c} =
100~h^{-1}~{\rm kpc}$ and $r_{co} = 10 \times r_{c}$ such that the
radial profile resembles the profiles of nearby clusters and that we
have optimized the cluster detections relative to the spurious
detection rates (Paper I). $\phi(m-m^{*})$ is the differential
Schechter luminosity function with $\alpha = -1.1$ and $M^{*} = -21.0$
and $-21.9$ in the \v4 and \i4 bands, respectively; $b(m)$ is the
background galaxy counts. For the derivation of these filters and a
detailed explanation, see Paper I. We have used extensive simulations
in both Paper I and this paper to quantify the selection bias due to
the functional form of the matched filter.  We find that this
selection bias has a minimal effect on the resulting distribution of
profiles.  The matched filter algorithm does a good job at preserving
the true distribution of profile shapes for a broad range of cluster
profile parameters.

Candidate clusters are detected individually in each band and then
matched with each other to locate those systems which are detected in
both bands. The catalog consists of 79 clusters of galaxies detected
with estimated redshifts between $z \sim 0.2$ and 1.2; the uncertainty
in the estimated redshift ($z_{est}$) is $\sigma_{z_{est}} \simless
0.2$ (see Paper I; Lubin 1995).  87\% of the cluster candidates are
matched detections (detected in both the \v4 and \i4 bands).  The
amplitude of the matched filter provides an estimate of the cluster
richness. This filter richness ($\Lambda_{cl}$) is a measure of the
effective number of $L^{*}$ galaxies in the cluster (see \S 4.2.2 of
Paper I; Lubin 1995). Through Monte-Carlo simulations, we can
statistically determine the relation between $\Lambda_{cl}$ and the
actual cluster richness as determined by the specification of Abell
(1958).  This relation is dependent on profile shape; as the cluster
profile slope steepens, a given $\Lambda_{cl}$ value corresponds to a
{\it lower} richness class. For a cluster which has a surface density
profile of approximately $r^{-1}$ (the assumed profile of the radial
filter), $\Lambda_{cl} \simgreat 40$ corresponds to Abell ${\rm R} \ge
1$ (Paper I).  Therefore, in the analysis which follows, we will
examine only those clusters with $\Lambda_{cl} \ge 40$.  The cluster
sample that will be the subject of this paper consists of 57 clusters
in the \v4 band and 50 clusters in the \i4 band.  Table 1 lists the ID
number of these clusters (according to the convention of Paper I).

\section{Individual Cluster Profiles}

\subsection{The Profile Fittings}

We begin by examining the individual profiles in the \v4 and \i4
bands.  We have an estimate of the cluster redshift ($z_{est}$) and
the cluster center from the matched filter detection algorithm.  The
cluster center is taken to be the position of the {\it peak} cluster
signal for each detection (see Paper I). Surface density profiles of
each cluster candidate are computed by counting the number of galaxies
in successive annuli around the center of each cluster.  We bin out to
a given {\it physical} radius of $1.0~h^{-1}$ Mpc (assuming
$\Omega_{o} = 1, H_{o} = 100~h~{\rm km~s^{-1}~Mpc^{-1}}, {\rm and}~h =
0.75$) at the estimated redshift of the cluster. Twenty bins of equal
radial extent are used. The uncertainty in each bin is calculated from
Poisson statistics. An annulus of radii $1.0 < r \le 1.5~h^{-1}~{\rm
Mpc}$ around each cluster is used to measure directly the background
level for each cluster. The measured background varies $\sim 10-15 \%$
from cluster to cluster and field to field. As an example, at a
redshift of $z \sim 0.5$, the number of cluster members (out to a
radius of $1.0~h^{-1}~{\rm Mpc}$) in a ${\rm R} = 1$ cluster is $\sim
150$ galaxies or 10 -- 20\% of the total number of galaxies in that
region.

We quantify the slope at large radii (described by a power-law
exponent $\alpha$) and the cluster core radius ($r_{c}$) of the
cluster profile by fitting each resulting background-subtracted
surface density profile to a King model (King 1962; Cavaliere \&
Fusco--Femiano 1976; Sarazin \& Bahcall 1977) :

\begin{equation}
S(r) = S_{o}~\left[1 + {\left({r \over r_{c}}\right)}^{2}\right]^{{-3
\beta + 1}\over 2}
\end{equation}

\noindent where $S_{o}$ and $r_{c}$ are the normalization and core
radius of the King model, respectively.  $\beta$ parameterizes the
density fall-off at large radii; that is, the surface density profile
falls off as $r^{-3 \beta + 1} \equiv r^{-\alpha}$.

Fig. 1 shows the distribution of power-law exponent ($\alpha$) and
core radii ($r_{c}$) for those cluster profiles which had acceptable
fits ($\chi^{2}/\nu \le 1.5; \nu = 17$).  The average $1\sigma$ single
parameter errors (Avni 1976) in the fit parameters of an {\it
individual} cluster profile are ${\Delta\alpha/\alpha} =
^{+16\%}_{-12\%}$ and ${\Delta r_{c}/r_{c}} = ^{+100\%}_{-60\%}$ ({\it
e.g.,} see Table 3).  We are able to constrain the slope of the
surface density profile much more accurately than the cluster core
radius (see also \S3.2).  As a consistency check, we have examined the
effect on the profile fittings by (1) calculating the cluster
background at a larger limiting radius; (2) changing the bin size (by
$\pm 50\%$); and (3) using the {\it weighted} center rather than the
peak center of each cluster (where the weighted center is defined as
the mean position as weighted by the amplitude of the matched filter
signal within the cluster detection; see also \S5); the cluster fit
parameters are all consistent within the measurement error with
$\Delta \alpha/\alpha \simless 10\%$ and $\Delta r_{c}/r_{c} \simless
50\%$.

The variation in the cluster profile parameters is large, as is the
sensitivity to the background level.  Therefore, it is difficult to
say anything quantitative on an individual basis (see also \S 4).
Table 2 lists the median values and ``$1 \sigma$'' confidence interval
for the {\it distribution} of $\alpha$ and $r_{c}$ in each band. We
define a $1 \sigma$ confidence interval around each parameter by
determining the points in the distribution such that the area in each
tail is 16\%. The average PDCS cluster has a surface density slope of
$\langle \alpha \rangle \approx 1.4$ and core radius of $\langle r_{c}
\rangle \approx 0.05~h^{-1}~{\rm Mpc}$. Note that the distribution of
core radii in the Palomar clusters indicates that approximately 20 --
30\% of the clusters have very small or essentially no core radii
(Fig.\ 1). The median values in the \v4 and \i4 bands are consistent
with each other (Table 2).

The observed range in $\alpha$ is completely consistent with a sample
of nearby, rich clusters ($\langle \alpha \rangle \approx 1.6$;
Bahcall 1977; see also Beers \& Tonry 1986; Oegerle \etal 1985;
Postman, Geller \& Huchra 1988), the galaxy--cluster cross correlation
function which represents the {\it average} net galaxy density profile
around rich clusters ($\langle \alpha \rangle \approx 1.2 - 1.4$;
Seldner \& Peebles 1977; Peebles 1980; Lilje \& Efstathiou 1988), and
measures of the mass (and light) distribution of intermediate redshift
clusters (Smail et al. 1995; Tyson \& Fischer 1995; Squires \etal
1995). The range of core radii of the PDCS clusters is also typical of
other well-studied nearby and intermediate redshift clusters. A study
of 15 rich, nearby clusters yields core radii of $r_{c} = 0.12 \pm
0.02~h^{-1}~{\rm Mpc}$ (Bahcall 1975; though a similar study of 12
clusters by Dressler 1978 indicate core radii larger by a factor of
$\sim 2$). Smail et al.\ (1995) examined the core radii of two
intermediate redshift, X-ray luminous clusters and found $r_{c} =
0.09^{+0.04}_{-0.03}~h^{-1}~{\rm Mpc}$ for 1455+22 ($z = 0.26$) and
$r_{c} \approx 0.16~h^{-1}~{\rm Mpc}$ for 0016+16 ($z = 0.55$), while
Molinari et al.\ (1994) found $r_{c} \approx 0.10 - 0.12~h^{-1}~{\rm
Mpc}$ for A3284 ($z = 0.15$), A3305 ($z = 0.16$), and A1942 ($z =
0.23$).

In Fig. 2, we show the cluster profiles for three matched PDCS cluster
candidates (cluster ID \# 003, 036 and 059; Paper I). In order to show
the diversity of the PDCS clusters, we have chosen clusters which
cover a range in estimated redshifts ($z_{est}$) and profile
shape. Table 3 lists the relevant parameters of the fittings to these
cluster profiles.

\subsection{Evaluating Selection Biases}

In order to estimate the measurement error and to demonstrate that the
matched filter does not significantly bias the profile distribution,
we have performed simple Monte-Carlo simulations. We estimate the
errors in our fitting procedure and evaluate the selection function by
running the cluster finding algorithm on artificial fields which
match, as closely as possible, the galaxy and cluster distributions of
the actual PDCS fields.  We reproduce our matching criteria in the
cluster detection method by creating ten pairs of galaxy fields, each
1 ${\rm deg^{2}}$, as viewed through the \v4 and \i4 passbands.  We
accomplished this by first modeling the actual background density --
magnitude function of regions of the PDCS fields which contain no
detected clusters. For each simulated field, we randomly distribute
the appropriate number of background galaxies in each magnitude bin
according to this function.  On top of this uniform field, we
superimpose 25 simulated clusters with the condition that the
separation between clusters must be greater than 500 arcsec. The
clusters are randomly given redshifts between 0.2 and 1.2 (in $z$
increments of 0.1) and richnesses of ${\rm R} \ge 0$.  The only
restriction placed on cluster richness is that clusters at $z \le 0.5$
cannot be richer than richness class 2 and clusters at $z \ge 0.6$
cannot be poorer than richness class 1. We have chosen for the cluster
luminosity function the same Schechter function as the matched filter;
however, we assume that half of the clusters contain spirals (and
apply a Scd spiral k--correction) and half contain ellipticals (and
apply an elliptical k--correction).  The k--correction will affect the
redshift estimate of the cluster ($\sigma_{z_{est}} \simless 0.2$) but
not the richness estimate $\Lambda_{cl}$ (Lubin 1995; Paper I).

Firstly, we estimate the {\it measurement} error inherent in our
fitting procedure by simulating a population of clusters with a {\it
unique} surface density slope ($\alpha$) and core radius ($r_{c}$). We
choose the average profile parameters observed in the PDCS : $\alpha =
1.4$ and $r_{c} = 0.05~h^{-1}~{\rm Mpc}$. The cluster galaxies are
distributed in an azimuthally symmetric manner according to the
appropriate King profile (Eq.\ 3). We apply the same detection
criteria and richness cut ($\Lambda_{cl} \ge 40$) to the simulated
clusters as for the PDCS clusters. For each simulated cluster, a
surface density profile is created and then fit to the King model in
the manner described in \S3.1. The resulting distributions of fit
parameters for 20 simulated clusters are shown in Fig.\ 3.  The median
values and their $1 \sigma$ errors (as defined in \S3.1) of these
distributions are listed in Table 4. The {\it median} values of the
distributions of fit parameters reproduce well the actual input
parameters of the simulated clusters; however, the distributions are
artifically broadened by $\sim 10\%$ in $\alpha$ and $\sim 50\%$ in
$r_{c}$. These errors are due to uncertainties, most notably
background subtraction, associated with the fitting procedure; in
addition, errors in the estimated redshift contribute to the broad
width of the $r_{c}$ distribution. The measurement uncertainties
associated with this particular cluster population are typical of the
errors for cluster populations with other combinations of $\alpha$ and
$r_{c}$.

Secondly, we examine the selection bias of the radial filter ($\alpha
= 1; r_{c} = 0.1~h^{-1}~{\rm Mpc}$) by exploring a broad range in
cluster shape through two profile distributions.  In the first set of
simulations, we generate a Gaussian distribution of profile shapes
which is reasonably close to that observed in the PDCS (Table 2); the
simulated clusters have a surface density slope distribution of
$\alpha = 1.4 \pm 0.6$ and a core radius distribution of $r_{c} = 0.05
\pm 0.05~h^{-1}~{\rm Mpc}$. In the second set of simulations, we
generate a Gaussian distribution of profile shapes which is very
different from the PDCS, $\alpha = 2.5 \pm 0.5$ and $r_{c} = 0.10 \pm
0.05 ~h^{-1}~{\rm Mpc}$. As above, the cluster galaxies are
distributed in an azimuthally symmetric manner according to the
appropriate King profile (Eq.\ 3); the same detection criteria and
richness cut ($\Lambda_{cl} \ge 40$) are applied. Fig.\ 4 shows the
input parameter distributions (shaded histograms) of the simulated
clusters and the output parameter distributions (solid line
histograms) of those clusters detected by the algorithm. There appears
to be {\it no} significant selection bias in either of the two
simulated cluster populations due to the matched filter; we reproduce
the input distributions of $\alpha$ and $r_{c}$ over this broad range
of cluster parameters extremely well. As noted in Paper I, clusters
with steeper profiles yield the strongest matched filter signal. This
will affect the cluster detection probability, though the exact effect
is also dependent on the cluster core radius and the cluster
richness. However, we have shown with the simulations of this section
that, for reasonable choices of the input cluster distribution (that
is, distributions with standard deviations typical of that observed in
the PDCS), the matched filter does not significantly bias the detected
(output) parameter distribution.

We have quantified both the measurement error associated with the
profile fitting procedure, as well as the selection bias associated
with the functional form of the matched filter. The observed median
parameters of the PDCS profile distributions appear to provide an
accurate indication of the true parent population of cluster shapes.

%We compare the observed PDCS profile distributions with the simulated
%cluster population ($\alpha = 1.4 \pm 0.6$; $r_{c} = 0.10 \pm
%0.05~h^{-1}~{\rm Mpc}$) by plotting in Fig. 1 the resulting $\alpha$
%and $r_{c}$ distributions for the simulated clusters detected by the
%matched filter algorithm (solid line histograms).  As expected, the
%simulated and the PDCS curves are not statistically different as
%measured through the Kolmogorov--Smirnov (KS) test, i.e. the null
%hypothesis that the actual and simulated distributions are drawn from
%the same parent population can only be rejected at confidence levels
%ranging between 28\% ($\alpha$; \v4 band) and 75\% ($r_{c}$; \i4
%band). The observed PDCS profile distributions should, thus, provide
%an accurate indication of the true parent distribution of cluster
%shapes.

\subsection{Central Density -- Richness Comparison}

We can also estimate the slope of the cluster surface density profiles
by examining the relation between richness ($N_r$) and central density
($N_o$) of the individual clusters. Similarly to Abell (1958), central
density and richness are defined as the number of member galaxies
(above background) that are brighter than $m_{3} + 2^{m}$ (where
$m_{3}$ is the magnitude of the third brightest galaxy) and located
within a projected radius of $0.25~h^{-1}~{\rm Mpc}$ and
$1.0~h^{-1}~{\rm Mpc}$, respectively (see \S 4.2.3 of Paper I for
details).  Fig. 5 shows $N_{r}$ versus $N_{o}$ for the PDCS clusters
in the two passbands (top panels).  The median values of $N_{r}/N_{o}$
are $2.5^{+1.0}_{-1.0}$ (\v4 band) and $2.7^{+1.4}_{-0.9}$ (\i4 band).
The standard deviation in these distributions represent $1 \sigma$
confidence intervals as defined in \S 3.1.

The expected relation between richness and central density for the
simulated cluster population described in \S 3.2 ($\alpha = 1.4 \pm
0.6$ and $r_{c} = 0.05 \pm 0.05 ~h^{-1}~{\rm Mpc}$) is shown in the
bottom panels of Fig. 5.  The median values of $N_{r}/N_{o}$ are
$2.4^{+1.1}_{-1.1}$ (\v4 band) and $2.7^{+1.0}_{-1.0}$ (\i4 band),
consistent with those of the PDCS. We note that the relation between
$N_{r}$ and $N_{o}$ for a population of clusters with a much {\it
steeper} surface density profile ({\it e.g.,} $r^{-2}$; $\langle
N_{r}/N_{o} \rangle \approx 1.6$) is inconsistent with that observed
in the PDCS; the probability, as measured by the KS test, that the
PDCS distribution is drawn from the same parent population as the
$N_{r}/N_{o}$ distribution of an $r^{-2}$ cluster population is
$\simless 10^{-12}\%$.

\section{Cluster Profile Evolution}

In an effort to improve our constraints on the cluster profiles, we
explore the average profile shape, specifically the surface density
slope $\alpha$, as a function of redshift by creating global
composites of our sample of PDCS clusters.  Because we exclusively use
estimated redshifts (observed redshifts are presently available for
fewer than 10 PDCS clusters) whose uncertainty are estimated at
$\sigma_{z_{est}} \simless 0.2$ (see \S 4.2.1 of Paper I), we sort
each cluster by its estimated redshift ($z_{est}$) in \v4 and \i4
bands into three broad redshift ranges : (1) $0.2 \le z_{est} \le
0.4$, (2) $0.5 \le z_{est} \le 0.7$, and (3) $0.8 \le z_{est} \le
1.2$. We examine the cluster profile at $r \ge 0.10~h^{-1}~{\rm Mpc}$
by creating a composite cluster profile.  We logarithmically bin
galaxies in the radial range $0.1 \le r \le 1.0~h^{-1}~{\rm Mpc}$
around each cluster center with a bin width of 0.1 in log $r$ (i.e.
10 bins).  The background level determined for each individual cluster
profile (from an outer annulus; see \S3.1) is first subtracted. We
stack the cluster profiles in each redshift interval according to
their physical radius ($0.1 \le r \le 1.0~h^{-1}~{\rm Mpc}$). The
resulting composite cluster profiles are then fit to a power-law of
the form $S(r) = S_{o} r^{-\alpha}$.

\subsection{Composite Cluster Profiles}

The composite profiles of the PDCS in the \v4 and \i4 bands are shown
in Fig. 6.  The redshift range is indicated in the upper right-hand
corner of each panel. The error in the mean value is determined from
the variance in each radius bin.  The solid line indicates the
best-fit power-law function.  Table 5 lists the number of cluster
profiles used to make the composite, the best-fit power-law slopes
$\alpha$, and their $1 \sigma$ single parameter errors. These errors
are calculated according to the procedure described by Avni (1976) who
presents the correct ``minimum $\chi^{2}$'' method for calculating
confidence limits in a fitting with one ``interesting parameter''
(i.e. $\alpha$). The slopes of the composite profiles are consistent
with the results of the individual fittings (Table 6).  The surface
density slopes of the most distant clusters ($z_{est} \ge 0.8$) are
slightly steeper than their intermediate redshift counterparts.  This
may simply reflect the larger uncertainty in the fitted parameters
when co-adding such a small number of clusters (9 in \v4 band and 5 in
\i4 band in this redshift interval). Such a large dispersion is also
seen in the simulations when co-adding a similar number of high
redshift clusters. In addition, it may be the result of a bias due to
the inherent difficulty in subtracting an appropriate background for
clusters at very large redshift. Nonetheless, some evolution in the
cluster profile is expected for certain cosmologies (see \S 6).

Since richer clusters are believed to have formed at earlier epochs
(Peebles 1993), we may expect to see a variation in profile shape with
richness. We, therefore, examine the composite cluster profiles of a
richer subset ($\Lambda_{cl} \ge 70$) of the PDCS clusters.  We find
that, within the observational uncertainty, there is no difference
between the power-law slopes of the composites of this richer
population and the composite profiles presented in Fig.\ 6.

\subsection{Evaluating Selection Biases}

In order to demonstrate that the composite profiles are not being
biased by a few clusters, we performed a parametric bootstrap analysis
(see Kendall \& Stuart 1967).  This technique involves randomly
drawing a new sample (of equal number) from the original cluster
sample. From each new sample, we create a composite cluster profile as
described above and fit the power-law function.  This procedure was
performed on 50 resampled datasets. The average (and standard
deviation) of the resulting $\alpha$ values are completely consistent
(within $1 \sigma$) with the fits to the original true composite in all
three redshift intervals (Table 5).

We confirm that our technique for creating cluster composites is not
biased due to either the method of background subtraction or the
functional form of the matched filter by performing two consistency
checks.  Firstly, we make composite cluster profiles from regions of
the PDCS fields which contain {\it no} detected clusters.  For each
``blank'' region, we randomly assign an estimated redshift drawn from
the actual estimated redshift distribution. We then bin the galaxies,
subtract the background, and stack the profiles in the same manner
used to create the PDCS composites.  The resulting composites are
shown in Fig. 7.  All profiles are consistent with zero, indicating
that our method of background subtraction is reasonably accurate.

Secondly, we use the simulations described in \S 3.2 to examine
whether our cluster composites accurately reflect the mean cluster
profile as a function of redshift.  In an identical manner to the
PDCS, we have created cluster composites from those clusters detected
in 10 simulated fields.  We have examined the simulated cluster
population which closely resembles the PDCS clusters ($\alpha = 1.4
\pm 0.6$ and $r_{c} = 0.05 \pm 0.05 ~h^{-1}~{\rm Mpc}$). The results
are shown in Fig.\ 8.  The best-fit $\alpha$ parameters are listed in
Table 6. The resulting composite profiles of the simulated clusters
are consistent with the mean slope in each redshift interval and in
each band.

We have shown that our actual cluster composites (Fig. 6) are a
reliable representation of the global cluster population and are not
significantly biased, as a function of redshift, by the intrinsic
nature of the matched filter.

\section{The Cluster Morphologies}

Observational studies of nearby clusters have shown that a significant
portion of clusters exhibit evidence of ellipticity, multiple
components, and other forms of substructure ({\it e.g.,} Geller \&
Beers 1982; Dressler \& Schectman 1988; West \& Bothun 1990; Jones \&
Forman 1992; Mohr et al.\ 1995; Buote \& Canizares 1995; de Theije,
Katgart \& van Kampen 1995). Presently, measurements such as the axial
ratio, centroid shift, and ellipticity have been used to quantify the
cluster ``morphology'' (the degree of asymmetry on large scales) in
the cluster surface density and X-ray contours (Mohr, Fabricant \&
Geller 1993; Mohr et al.\ 1995; Buote \& Canizares 1995; de Theije et
al.\ 1995).

We take a similar approach by examining the degree of symmetry in the
PDCS clusters through the ``filtered'' images of the PDCS fields,
i.e.\ the amplitude of the matched filter signal as a function of
position within a given cluster detection.  At each position, this
signal measures how accurately a cluster centered at that position
matches the filter.  Since the filter is an axisymmetric,
monotonically decreasing function of radius, the amplitude as a
function of position is effectively a map of the structure in the
cluster candidate.  The first two columns of Fig.\ 9 show the contour
plots of sample PDCS clusters in each of the three redshift intervals
(top to bottom). Adjacent panels show the contours of each cluster in
the \v4 and \i4 band, respectively.  The side of each panel
corresponds to $2~h^{-1}~{\rm Mpc}$ at the estimated redshift of the
cluster.  The contours represent lines of constant amplitude of the
matched filter signal.

We compare these contour maps with the azimuthally symmetric,
centrally concentrated clusters of our simulations (\S 3.2). The last
two columns of Fig.\ 9 show the contour levels of simulated clusters
at the same estimated redshifts as the PDCS clusters. The simulated
symmetric clusters may be marginally rounder than the PDCS
clusters. We quantify any difference by measuring the offset between
the {\it peak} center and the {\it weighted} center of the cluster.
The peak center (used in the profile analysis) is defined as the
position of the maximum signal in a given cluster detection; the
weighted center is defined as the mean position as weighted by the
amplitude of the matched filter signal within the cluster detection.
The magnitude of this offset is a measure of the degree of asymmetry
of the clusters.  Fig. 10 shows the distribution of the offset (in
$h^{-1}~{\rm kpc}$) for the PDCS clusters (shaded histograms) and for
the azimuthally symmetric simulated clusters (solid line histograms).
Using the KS test, the null hypothesis that the offset distributions
are drawn from the same parent cluster population can be rejected at a
99.9\% confidence level for the \v4 band and at a greater than
99.9999\% confidence level for the \i4 band, indicating that the
typical PDCS cluster is not azimuthally symmetric.  That is, the PDCS
clusters are inconsistent with an azimuthally symmetric cluster
population with its characteristic $\alpha$ and core radius
$r_{c}$. Significant structure on large scales, including ellipticity
and bi-modality, in both mass, light, and X-rays have been previously
found by Smail et al. (1995), Tyson \& Fischer (1995) and Squires
\etal (1995) in the intermediate redshift clusters 1455+22 ($z =
0.26$), 0016+16 ($z = 0.55$), Abell 1689 ($z = 0.18$), and Abell 2218
($z = 0.175$).

\section{Comparison to Structure Formation Simulations}

Recently, Crone et al. (1994) and Jing \etal (1995) have used high
resolution N-body simulations to examine the dependence of cosmology
on the cluster mass density profile and morphology.  The PDCS provides
a large statistical sample to which we can compare the results of
these cosmological simulations.  Crone et al. (1994) examine the
dependence of the cluster profile on both cosmology and initial
density field by using a scale-free initial power spectra $P(k)
\propto k^{n}$ with initial spectral indices $n = -2, -1$ and 0 for
four cosmologies : Einstein deSitter ($\Omega_{o} = 1$), open
($\Omega_{o} = 0.2$ and 0.1), and flat, low-density ($\Omega_{o} =
0.2$, $\Lambda_{o} = 0.8$).  Jing et al. (1995) examined the Standard
CDM model and six low-density CDM models with and without a
cosmological constant.  The resulting average mass density profiles
are all well-fit by a power-law $\rho(r) \propto r^{-\gamma}$ for
radii greater than $\sim 0.2~h^{-1}~{\rm Mpc}$ (or local density
contrasts between 100 and 3000).  There exists a clear trend toward
steeper slopes with decreasing $\Omega_{o}$. Crone et al. (1994) also
found a correlation between steepening slopes and increasing spectral
index $n$ in their initial power spectrum. In addition, Jing et
al. (1995) find that, for a given $\Omega_{o} (< 1)$, clusters in the
flat model ($\Lambda_{o} = 1 - \Omega_{o}$) have flatter mass density
profiles than in the corresponding open model.  A composite of these
results are presented in Fig. 11 where we plot the slope of the {\it
surface} mass density profile ($\alpha = \gamma - 1$) versus
$\Omega_{o}$ for simulated clusters at $z = 0$.  Flat cosmologies are
indicated by closed circles, while open cosmologies are indicated by
open circles.

We would now like to compare the composite surface density profiles of
the PDCS clusters with the cluster simulations; however, the
simulations model the total mass distribution, whereas we examine the
galaxy distribution.  It is now possible to map the total mass density
distribution as a function of radius through observations in X-rays
and through gravitational lensing of rich clusters of galaxies ({\it
e.g.,} Henry, Briel \& Nulsen 1994; Smail \etal 1995; Tyson
\& Fischer 1995; Squires \etal 1995).  These observations reveal that
there is excellent agreement between the projected structure of the
galaxy, gas, and dark matter distributions on scales larger than $\sim
100~h^{-1}~{\rm kpc}$.  Therefore, a comparison between the composite
surface density profiles of the PDCS clusters with those expected from
the large scale simulations should allow reasonable constraints to be
placed on the mass profiles.  As discussed in Tyson \& Fischer (1995),
our mean projected density slope of $\langle \alpha
\rangle \approx 1.4$ is intermediate between standard CDM and an
$\Omega_{o} = 0.35$ open CDM or $\Lambda +$ CDM simulated profiles but
appears to be a better match to the latter two. In Fig. 11, we
indicate by two dotted lines the range of power-law slopes that are
obtained from the composite PDCS cluster profiles over the redshift
range $0.2 \simless z \simless 1.2$ (Table 5 and Fig. 6).  The
$\Omega_{o} = 1$ profiles are indeed on the low end of our observed
$\alpha$ distribution; however, this is far from conclusive.

Up to this point there has been little examination of the evolution of
cluster profile shape as a function of redshift because the cluster
samples at intermediate to high redshift have been sparse.  The
Palomar survey provides the first statistically complete cluster
survey over a large redshift range.  The cluster composites may
indicate that the profiles get {\it steeper} with increasing redshift.
Such a trend in the mean slope is most prominent in the \i4 band
though the trend is not significant given the large observational
uncertainty. However, such an effect may also be observed in simulated
clusters from high-resolution (20 kpc) N body simulations (Xu 1995).
Because clusters form much later in an $\Omega_{o} = 1$ universe,
standard CDM shows the largest evolution in the cluster profile. The
slope decreases from $\alpha = 1.51 \pm 0.27$ ($z = 1$) to $\alpha =
1.29 \pm 0.19$ ($z = 0$). The shallowing of the power-law slope at
radii greater than $\sim 0.2 ~h^{-1}~{\rm Mpc}$ is largely the result
of an {\it increase} in the cluster core radius due to continued
merging between $z = 1$ and $z = 0$ (Xu 1995; Pen 1995).  Profiles of
clusters in an open ($\Omega_{o} = 0.35$) CDM shows no detectable
evolution with $\alpha = 1.38 \pm 0.38$ ($z = 1$) and $\alpha = 1.38
\pm 0.19$ ($z = 0$). $\Omega_{o} + \Lambda$ models yield intermediate
results.  Other non-standard cosmologies, such as mixed dark matter
models, also show a steepening of the cluster profile with redshift
(Walter \& Klypin 1995). At present, with little redshift information,
it is not possible to use the observed variation to rule out any
cosmological model.

The cluster morphology, or the degree of symmetry, has also been used
to delineate between cosmological models (Evrard et al.\ 1993; Mohr et
al.\ 1995; Jing et al. 1995).  N-body simulations and hydrodynamic
simulations show that the surface density and X-ray image contours of
clusters in low-density CDM models ({\it e.g.,} $\Omega_{o} = 0.2$)
are much more regular and centrally concentrated that those in an
Einstein-de Sitter model ($\Omega_{o} = 1$). Clusters in low
$\Omega_{o}$ cosmologies are older and are, therefore, more relaxed
and symmetric than clusters in an $\Omega_{o} = 1$ universe. Mohr
\etal (1995) have used these results to argue in favor of a flat
universe, given the observed degree of asymmetry and structure in
present-day clusters.  Jing et al. (1995) found, however, that
low-$\Omega_{o}$ models with a cosmological constant $\Lambda$ ({\it
e.g.,} $\Omega_{o} = 0.3$ and $\Lambda = 0.7$) also produce a large
fraction of clusters with significant deviations from spherical
symmetry, though the density and X-ray image contours are still
rounder than the $\Omega_{o} = 1$ case.  However, this matter is still
controversial as Pen (1995) and Xu (1995) find little quantitative
difference in the structure characteristics of clusters formed in
these cosmological models.  We have used the contours of constant
amplitude of the matched filter signal to examine the degree of
symmetry in the PDCS clusters.  We do find that the PDCS clusters are
inconsistent with a population of azimuthally symmetric simulated
clusters (Fig.\ 10).

\section{Summary}

We have examined the galaxy distributions of the richest clusters of
galaxies from the Palomar Distant Cluster Survey.  Through an analysis
of individual cluster profiles and composite profiles as a function of
redshift, we find that the typical Palomar cluster has a profile of
$r^{-1.4}$ at radii greater than $0.10~h^{-1}~{\rm Mpc}$ and a core
radius of $0.05~h^{-1}~{\rm Mpc}$. The distribution of core radii in
the Palomar clusters indicates that up to 30\% of the clusters have
very small or essentially no core radii. Using simple Monte-Carlo
simulations, we have shown that the median cluster profile parameters
appear to be an accurate representation of the actual cluster
population; the measurement errors associated with these parameters
are $\sim 10\%$ in the surface density slope $\alpha$ and $\sim 50\%$
in the cluster core radius $r_{c}$. The average profile parameters of
the PDCS clusters are consistent with measures of nearby clusters, as
well as other intermediate redshift clusters (Molinari et al.\ 1994;
Tyson \& Fischer 1995; Squires et al.\ 1995; Smail et al.\
1995). There may be an indication that the mean cluster profile
steepens with increasing redshift, though the observational
uncertainty is presently too large to be conclusive.  In addition, we
find that a large fraction of the PDCS clusters have a significant
degree of asymmetry.  The PDCS cluster population is inconsistent with
a population of circularly symmetric clusters of galaxies at a greater
than 99.9\% confidence level.

\vskip 0.7cm

We thank the anonymous referee for comments which greatly improved
this paper. We gratefully acknowledge the useful discussions with Neta
Bahcall, Jim Gunn, Robert Lupton, David Spergel, and Michael
Strauss. We also thank Sangeeta Malhotra for her library of SM macros,
Bob Rutledge for his incredible fitting program BFIT, and Ue-Li Pen
and Guohong Xu for preliminary results of their cosmological
simulations. This work is supported in part by NASA contract NGT-51295
(LML).

\vfill
\eject

\vfill
\eject

\centerline{\bf Figure Captions}

\newcounter{discnt}

\begin{list}
{{\bf Figure \arabic{discnt}} :}  {\usecounter{discnt}}

% figure 1
\item Distribution of surface density slopes ($\alpha$) and core
radii ($r_{c}$ in $h^{-1}~{\rm Mpc}$) from the fittings to the PDCS
cluster profiles.  The shaded and solid line histograms represent the
distribution of parameters from individual profile fittings of the
PDCS clusters for the \v4 and \i4 bands, respectively.  We show only
those fittings which have acceptable $\chi^{2}/\nu$ (see \S 3.1).

% figure 2 
\item King profile fittings to three matched cluster
candidates. Adjacent panels show the cluster profile in the \v4 and
\i4 bands, respectively.  The estimated redshift ($z_{est}$) and the
cluster ID number is indicated in the top right and left of each
panel, respectively. A solid line indicates the best-fit King model
(Eq.\ 3).

% figure 3
\item Distribution of surface density slopes ($\alpha$) and core
radii ($r_{c}$ in $h^{-1}~{\rm Mpc}$) from the fittings to the
profiles of a simulated cluster population with a discreet $\alpha$
and $r_{c}$ combination ($\alpha = 1.4$; $r_{c} = 0.05~h^{-1}~{\rm
Mpc}$). The median values accurately reproduce the actual profile
parameters, though the distributions are artificially broadened due to
inherent errors in the fitting process (see \S3.2).

%figure 4
\item The input (shaded histograms) and output (solid line histograms)
distributions of surface density slopes ($\alpha$; top panels) and
core radii ($r_{c}$ in $h^{-1}~{\rm Mpc}$; bottom panels) of the two
sets of simulated clusters (see \S 3.2). The input parameter
distributions are specified in the upper right-hand corner of each
panel. All distributions are shown separately for the
\v4 and \i4 bands.

%figure 5
\item Abell richness $N_{r}~(\le 1.0~h^{-1}~{\rm Mpc})$ versus central
density $N_{o}~(\le 0.25~h^{-1}~{\rm Mpc})$ for the PDCS clusters and
for simulated clusters (see \S 3.2) in the \v4 and \i4 bands.  The
black dots, open squares, and stars indicate clusters detected at
estimated redshifts of $0.2 \le z_{est} \le 0.4$, $0.5 \le z_{est} \le
0.7$, and $z_{est} \ge 0.8$, respectively. The expected relation
from the simulations is consistent with the observations.

% figure 6
\item Composite surface density profiles ($0.10 \le r \le
1.0~h^{-1}~{\rm Mpc}$) of the PDCS clusters.  The clusters have been
divided in three redshift bins based upon their estimated redshifts in
each band.  The individual background surface density has been
subtracted from each individual cluster profile before creating the
composite.  The variance in each bin is used to compute the error in
the mean value. The best-fit power-law functions are shown.

% figure 7
\item Composite surface density profiles
($0.1 \le r \le 1.0~h^{-1}~{\rm Mpc}$) of regions in the PDCS fields
which contain no detected clusters.  We have assumed estimated
redshifts such that we have roughly the same number of profiles in
each redshift bin as the real cluster composites (Fig. 6). The
constant background surface density has been subtracted from each
``profile'' before creating the composite. All profiles are consistent
with zero, indicating that our method of background subtraction is
reasonably accurate.

% figure 8
\item Composite surface density profiles
($0.1 \le r \le 1.0~h^{-1}~{\rm Mpc}$) of the simulated clusters.
The clusters have been divided in three redshift bins based upon their
estimated redshifts in each band.  The constant background surface
density has been subtracted from each individual cluster profile
before creating the composite.  The variance in each bin is used to
compute the error in the mean value. The best-fit power-law functions
are shown.

% figure 9
\item Contour maps of the amplitude of the matched filter signal in
both bands for one cluster in each of the three redshift intervals
(top panel $0.2 \le z_{est} \le 0.4$; middle panel $0.5 \le z_{est}
\le 0.7$; bottom panel $0.8 \le z_{est} \le 1.2$).  The first two
columns show actual PDCS clusters, and the last two columns show
azimuthally symmetric, simulated clusters. The side of each panel
corresponds to $2~h^{-1}~{\rm Mpc}$ at the estimated redshift of the
cluster.

% figure 10
\item Distribution of offsets (in $h^{-1}~{\rm kpc}$) between the peak
center (the position of the maximum signal of the matched filter) and
the weighted center.  The shaded histograms represent the distribution
of offsets of the PDCS clusters. The solid line histograms represent
the distribution of offsets of the azimuthally symmetric, simulated
clusters. The distributions are shown separately for the \v4 and \i4
bands. The actual data and the simulations are statistically different
(see \S 5).

% figure 11
\item The expected power-law slopes of the cluster {\it mass}
surface density profiles versus $\Omega_{o}$ from various cosmological
simulations (Crone et al. 1994; Jing et al. 1995).  Open and closed
circles indicate open and flat cosmologies, respectively, and
represent simulated clusters at $z = 0$.  Dashed lines indicate the
range of power-law slopes of the composite PDCS cluster {\it galaxy}
profiles over our full redshift interval $0.2 \simless z \simless 1.2$
(Table 5).

\end{list}

\vfill
\eject

% Table 1
\begin{table}
\begin{center}
Table 1 : The Sample of PDCS Clusters
\end{center}
\begin{center}
\begin{tabular}[t]{cl}
\hline
\hline
Band	&	PDCS Cluster ID Number	\\
\hline
\v4	&	001 002 003 004 006 008 010 011 012 015 \\
	&	016 017 018 019 020 021 022 023 024 025 \\
	&	027 030 031 033 034 035 036 037 038 039 \\
	&	041 042 044 047 048 049 050 051 052 053 \\
	&	054 055 056 057 059 062 063 065 066 067 \\
	& 	068 069 071 072 075 076 078 		\\
\hline
\i4	&	001 002 003 004 006 008 009 010 011 012 \\
	&	014 015 016 017 018 019 020 021 022 023 \\
	&	024 031 033 034 035 036 037 039 041 042 \\
	&	045 049 050 051 052 054 055 056 057 059 \\
	&	061 062 063 064 067 068 069 075 076 079	\\
\hline
\end{tabular}
\end{center}
\end{table}

% Table 2
\begin{table}
\begin{center}
Table 2 : Median Values of Surface Density Slope $\alpha$ and Core
Radius $r_{c}~(h^{-1}~{\rm Mpc})$ of the PDCS Clusters
\end{center}
\begin{center}
\begin{tabular}[t]{cccc}
\hline
\hline
Band	&	\#	&	$\alpha$	&$r_{c}$	\\
\hline
\v4	&	43& $1.36^{+1.12}_{-0.47}$& $0.05^{+0.12}_{-0.04}$	\\
\i4	& 	37& $1.35^{+1.07}_{-0.35}$& $0.05^{+0.10}_{-0.03}$	\\
\hline
\end{tabular}
\end{center}
\end{table}

% Table 3
\begin{table}
\begin{center}
Table 3 : Surface Density Slope $\alpha$ and Core Radius $r_{c}~(h^{-1}~{\rm
Mpc})$ of Three Matched PDCS Clusters
\end{center}
\begin{center}
\begin{tabular}[t]{cccccc}
\hline
\hline
	& PDCS	&\multicolumn{2}{c}{\v4}	&\multicolumn{2}{c}{\i4} \\
Panel	& ID \#	& $\alpha$& $r_{c}$ 	& $\alpha$ & $r_{c}$ \\
\hline
Top    & 036	& $1.72^{+0.39}_{-0.23}$& $0.03^{+0.04}_{-0.02}$ & $2.53^{+0.69}_{-0.42}$ &
$0.21^{+0.46}_{-0.11}$ \\
Middle & 003	& $1.79^{+0.66}_{-0.42}$& $0.35^{+0.05}_{-0.29}$ & $1.23^{+0.18}_{-0.13}$ &
$0.04^{+0.03}_{-0.02}$ \\
Bottom & 059	& $2.13^{+0.75}_{-0.39}$& $0.04^{+0.25}_{-0.02}$ & $2.42^{+0.51}_{-0.30}$ &
$0.04^{+0.04}_{-0.02}$ \\
\hline
\end{tabular}
\end{center}
\end{table}

% Table 4
\begin{table}
\begin{center}
Table 4 : Measurement Error Associated with Median Values of Surface
Density Slope $\alpha$ and Core Radius $r_{c}~(h^{-1}~{\rm Mpc})$ from
the Simulated Clusters
\end{center}
\begin{center}
\begin{tabular}[t]{cccc}
\hline
\hline
Band	&	$\alpha$	&$r_{c}$	\\
\hline
\v4	& $1.36^{+0.12}_{-0.23}$& $0.057^{+0.032}_{-0.015}$	\\
\i4	& $1.43^{+0.12}_{-0.15}$& $0.059^{+0.032}_{-0.027}$	\\
\hline
\end{tabular}
\end{center}
\end{table}

% Table 5
\begin{table}
\begin{center}
Table 5 : Power-Law Slopes of Composite Surface Density Profiles of the PDCS Clusters
\end{center}
\begin{center}
\begin{tabular}[t]{ccccc}
\hline
\hline
Redshift Interval&\multicolumn{2}{c}{\v4}	&\multicolumn{2}{c}{\i4} \\
	& \#		& $\alpha$ 	& \# 	&	$\alpha$ \\
\hline
$0.2 \le z_{est} \le 0.4$& 26 & $1.50^{+0.22}_{-0.34}$& 21 & $1.32^{+0.13}_{-0.18}$\\
$0.5 \le z_{est} \le 0.7$& 22 & $1.34^{+0.18}_{-0.27}$& 24 & $1.66^{+0.16}_{-0.24}$\\
$0.8 \le z_{est} \le 1.2$& 9  & $2.06^{+0.24}_{-0.43}$&  5 & $1.93^{+0.19}_{-0.24}$\\
\hline
\end{tabular}
\end{center}
\end{table}

% Table 6
\begin{table}
\begin{center}
Table 6 : Power-Law Slopes of Composite Surface Density Profiles of the Simulated Clusters
\end{center}
\begin{center}
\begin{tabular}[t]{ccccc}
\hline
\hline
Redshift Interval&\multicolumn{2}{c}{\v4}	&\multicolumn{2}{c}{\i4} \\
	& \#		& $\alpha$ 	& \# 	&	$\alpha$ \\
\hline
$0.2 \le z_{est} \le 0.4$& 29 & $1.46^{+0.15}_{-0.20}$& 29 & $1.47^{+0.15}_{-0.19}$\\
$0.5 \le z_{est} \le 0.7$& 66 & $1.49^{+0.15}_{-0.20}$& 71 & $1.45^{+0.12}_{-0.16}$\\
$0.8 \le z_{est} \le 1.2$& 45 & $1.37^{+0.25}_{-0.55}$& 53 & $1.44^{+0.15}_{-0.20}$\\
\hline
\end{tabular}
\end{center}
\end{table}

\vfill
\eject

% figure 1
\begin{figure}
\centerline{
\epsfysize=7.5in
\epsfbox{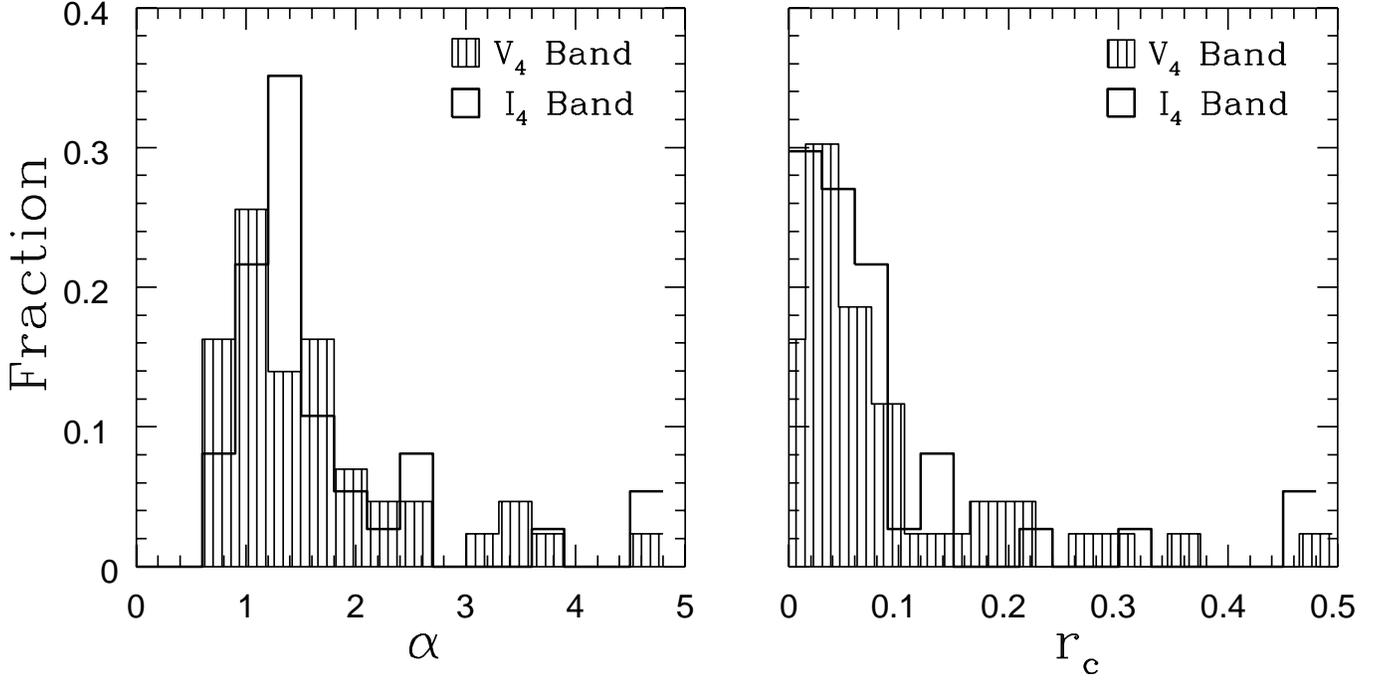}}
\caption{Distribution of surface density slopes ($\alpha$) and core
radii ($r_{c}$ in $h^{-1}~{\rm Mpc}$) from the fittings to the PDCS
cluster profiles.  The shaded and solid line histograms represent the
distribution of parameters from individual profile fittings of the
PDCS clusters for the \v4 and \i4 bands, respectively.  We show only
those fittings which have acceptable $\chi^{2}/\nu$ (see \S 3.1).}
\end{figure}

% figure 2 
\begin{figure}
\centerline{
\epsfysize=7.5in
\epsfbox{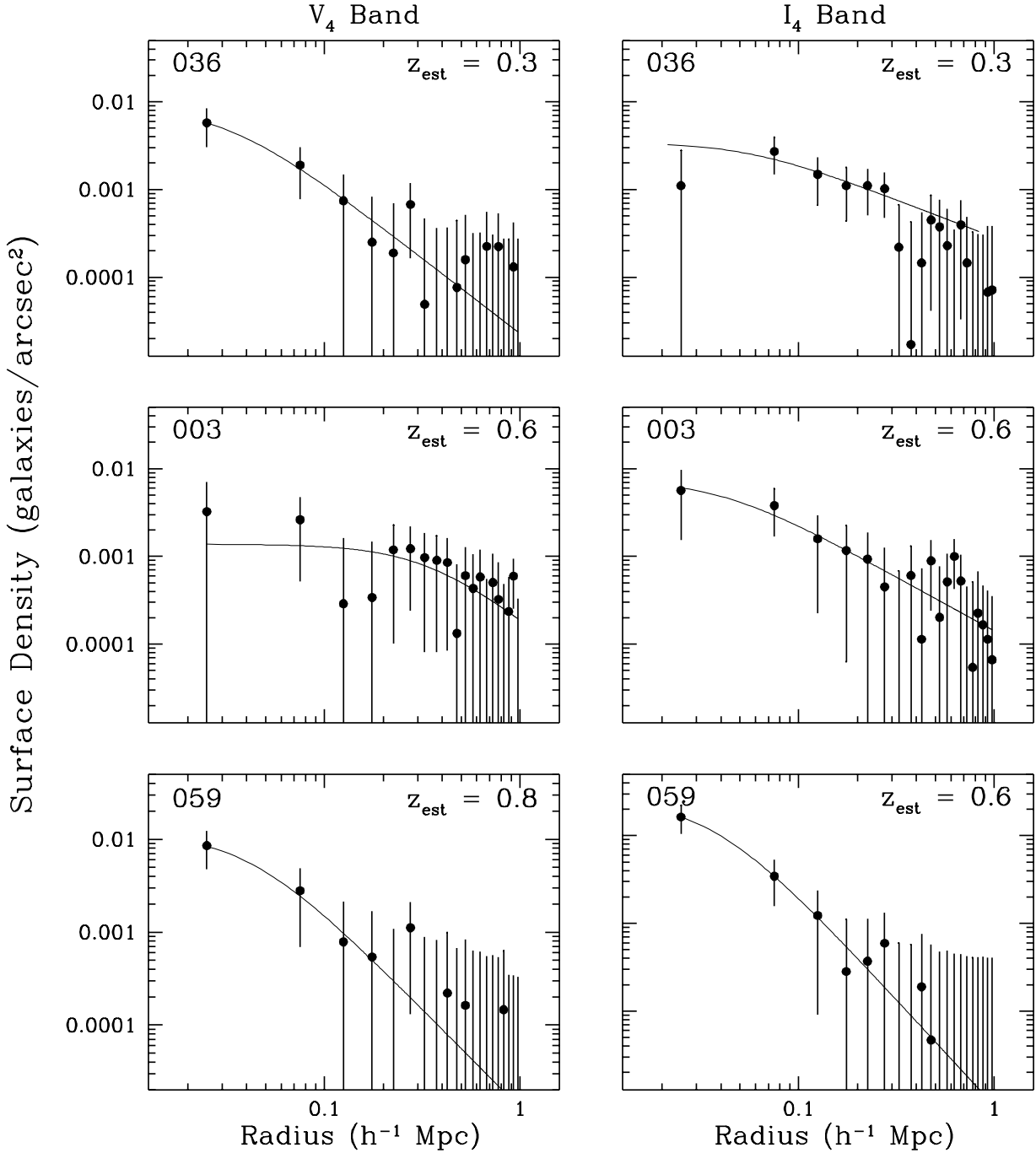}}
\baselineskip 14pt
\caption{King profile fittings to three matched cluster
candidates. Adjacent panels show the cluster profile in the \v4 and
\i4 bands, respectively.  The estimated redshift ($z_{est}$) and the
cluster ID number is indicated in the top right and left of each
panel, respectively. A solid line indicates the best-fit King model
(Eq.\ 3).}
\end{figure}

% figure 3
\begin{figure}
\centerline{
\epsfbox{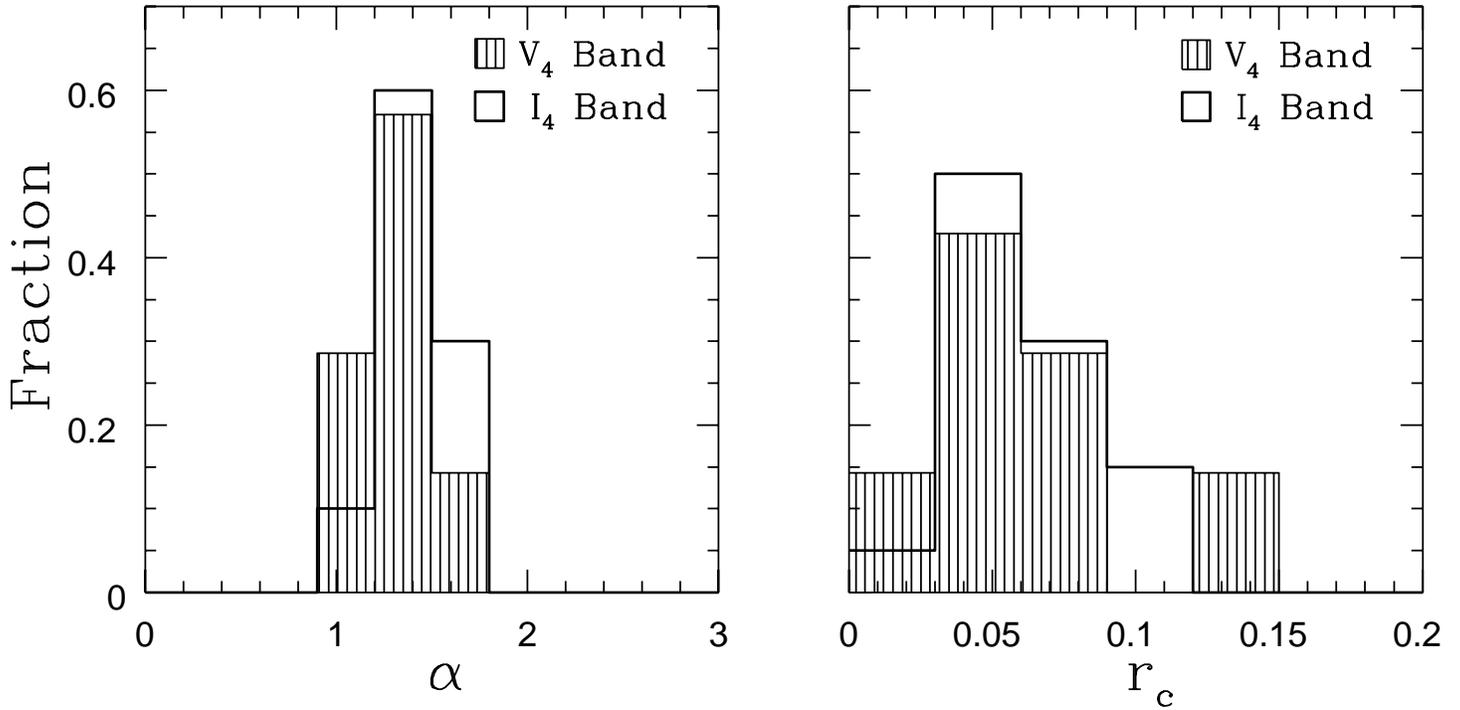}}
\baselineskip 14pt
\caption{Distribution of surface density slopes ($\alpha$) and core
radii ($r_{c}$ in $h^{-1}~{\rm Mpc}$) from the fittings to the
profiles of a simulated cluster population with a discreet $\alpha$
and $r_{c}$ combination ($\alpha = 1.4$; $r_{c} = 0.05~h^{-1}~{\rm
Mpc}$). The median values accurately reproduce the actual profile
parameters, though the distributions are artificially broadened due to
inherent errors in the fitting process (see \S3.2).}
\end{figure}

%figure 4
\begin{figure}
\centerline{
\epsfysize=7.5in
\epsfbox{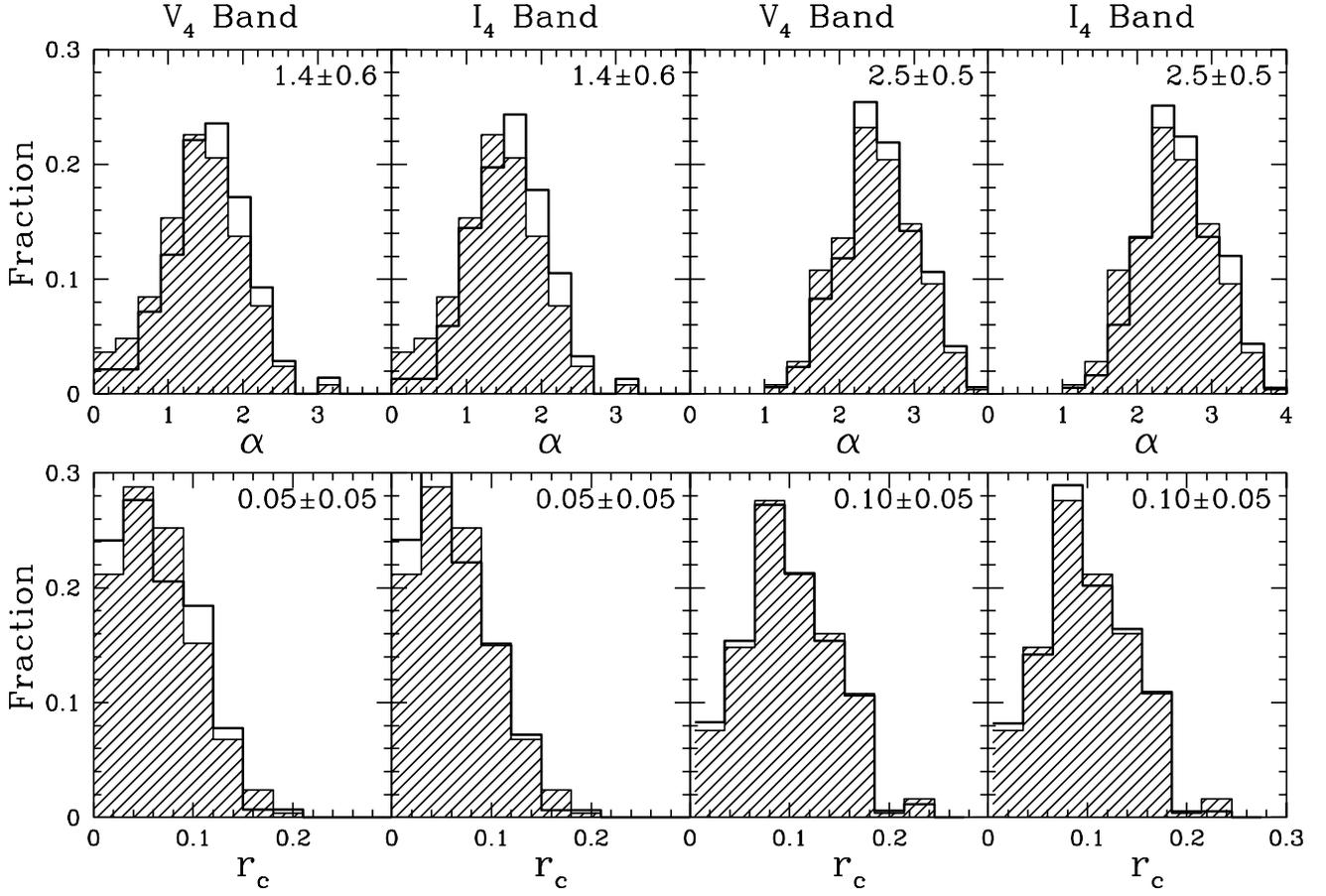}}
\caption{The input (shaded histograms) and output (solid line histograms)
distributions of surface density slopes ($\alpha$; top panels) and
core radii ($r_{c}$ in $h^{-1}~{\rm Mpc}$; bottom panels) of the two
sets of simulated clusters (see \S 3.2). The input parameter
distributions are specified in the upper right-hand corner of each
panel. All distributions are shown separately for the
\v4 and \i4 bands.}
\end{figure}

%figure 5
\begin{figure}
\centerline{\epsfbox{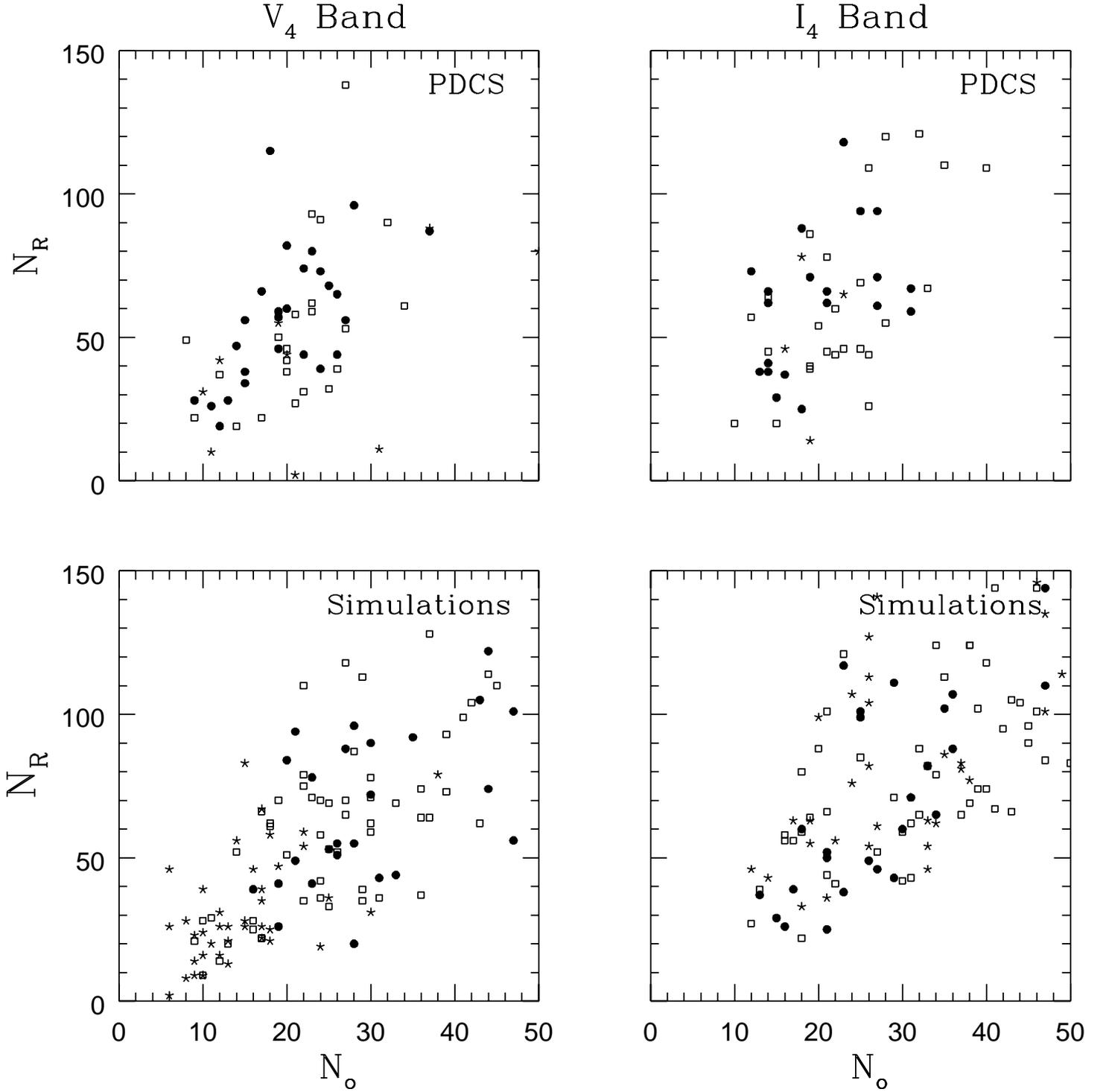}}
\caption{Abell richness $N_{r}~(\le 1.0~h^{-1}~{\rm Mpc})$ versus central
density $N_{o}~(\le 0.25~h^{-1}~{\rm Mpc})$ for the PDCS clusters and
for simulated clusters (see \S 3.2) in the \v4 and \i4 bands.  The
black dots, open squares, and stars indicate clusters detected at
estimated redshifts of $0.2 \le z_{est} \le 0.4$, $0.5 \le z_{est} \le
0.7$, and $z_{est} \ge 0.8$, respectively. The expected relation
from the simulations is consistent with the observations.}
\end{figure}

% figure 6
\begin{figure}
\centerline{
\epsfysize=7.5in
\epsfbox{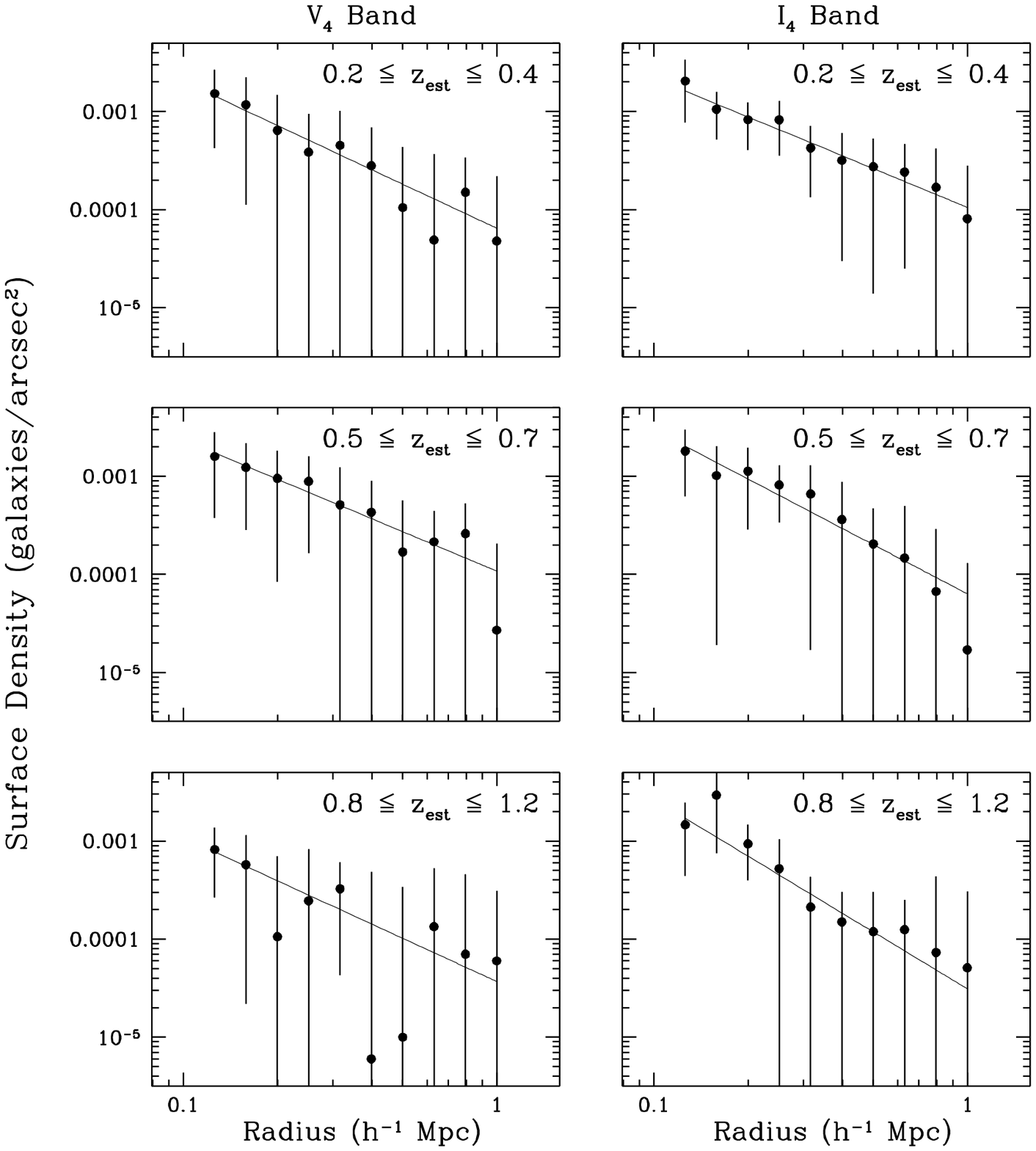}}
\baselineskip 14pt
\caption{Composite surface density profiles ($0.10 \le r \le
1.0~h^{-1}~{\rm Mpc}$) of the PDCS clusters.  The clusters have been
divided in three redshift bins based upon their estimated redshifts in
each band.  The individual background surface density has been
subtracted from each individual cluster profile before creating the
composite.  The variance in each bin is used to compute the error in
the mean value. The best-fit power-law functions are shown.}
\end{figure}

% figure 7
\begin{figure}
\centerline{
\epsfysize=7.5in
\epsfbox{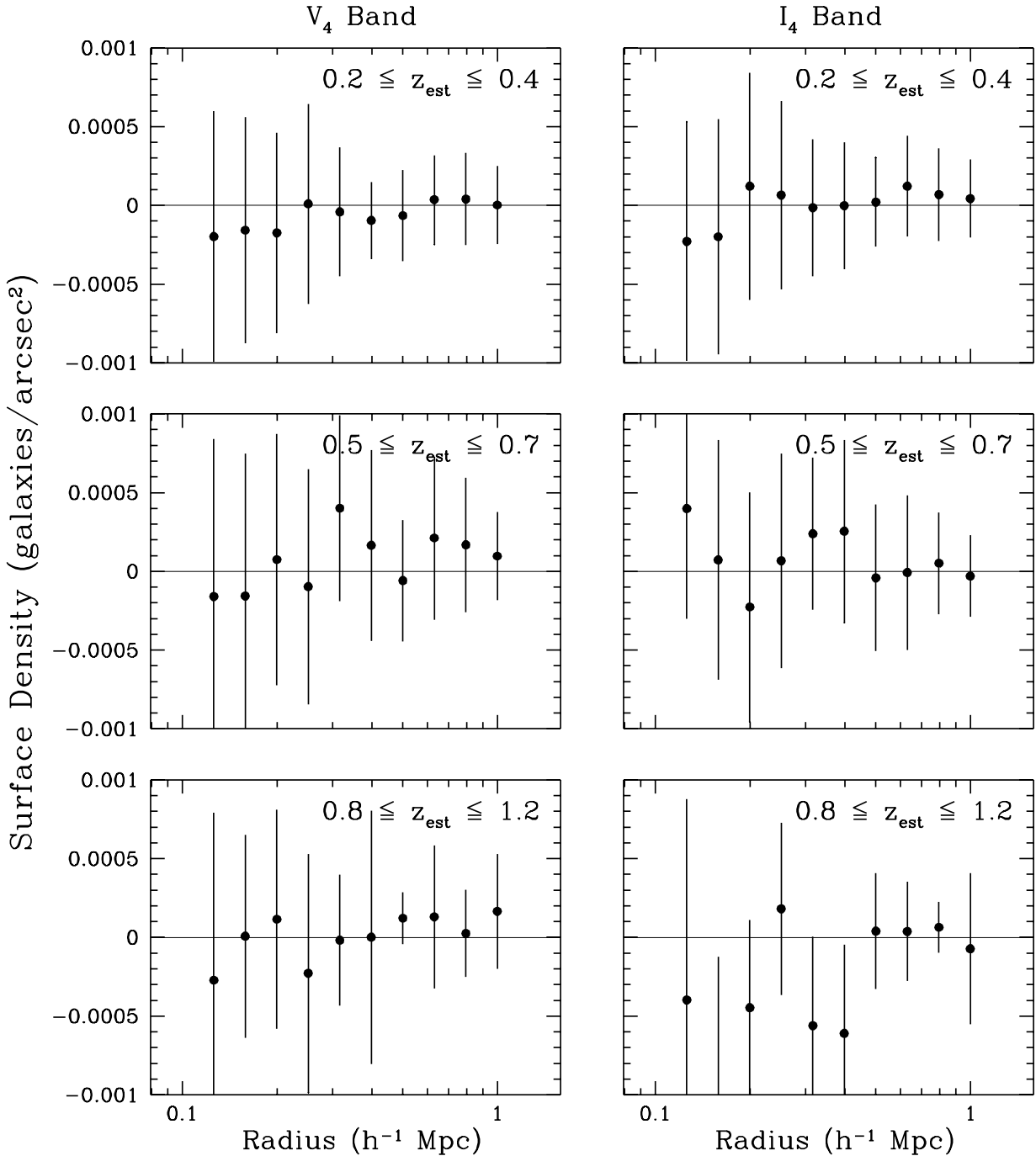}}
\caption{Composite surface density profiles
($0.1 \le r \le 1.0~h^{-1}~{\rm Mpc}$) of regions in the PDCS fields
which contain no detected clusters.  We have assumed estimated
redshifts such that we have roughly the same number of profiles in
each redshift bin as the real cluster composites (Fig. 6). The
constant background surface density has been subtracted from each
``profile'' before creating the composite. All profiles are consistent
with zero, indicating that our method of background subtraction is
reasonably accurate.}
\end{figure}

% figure 8
\begin{figure}
\centerline{
\epsfysize=7.5in
\epsfbox{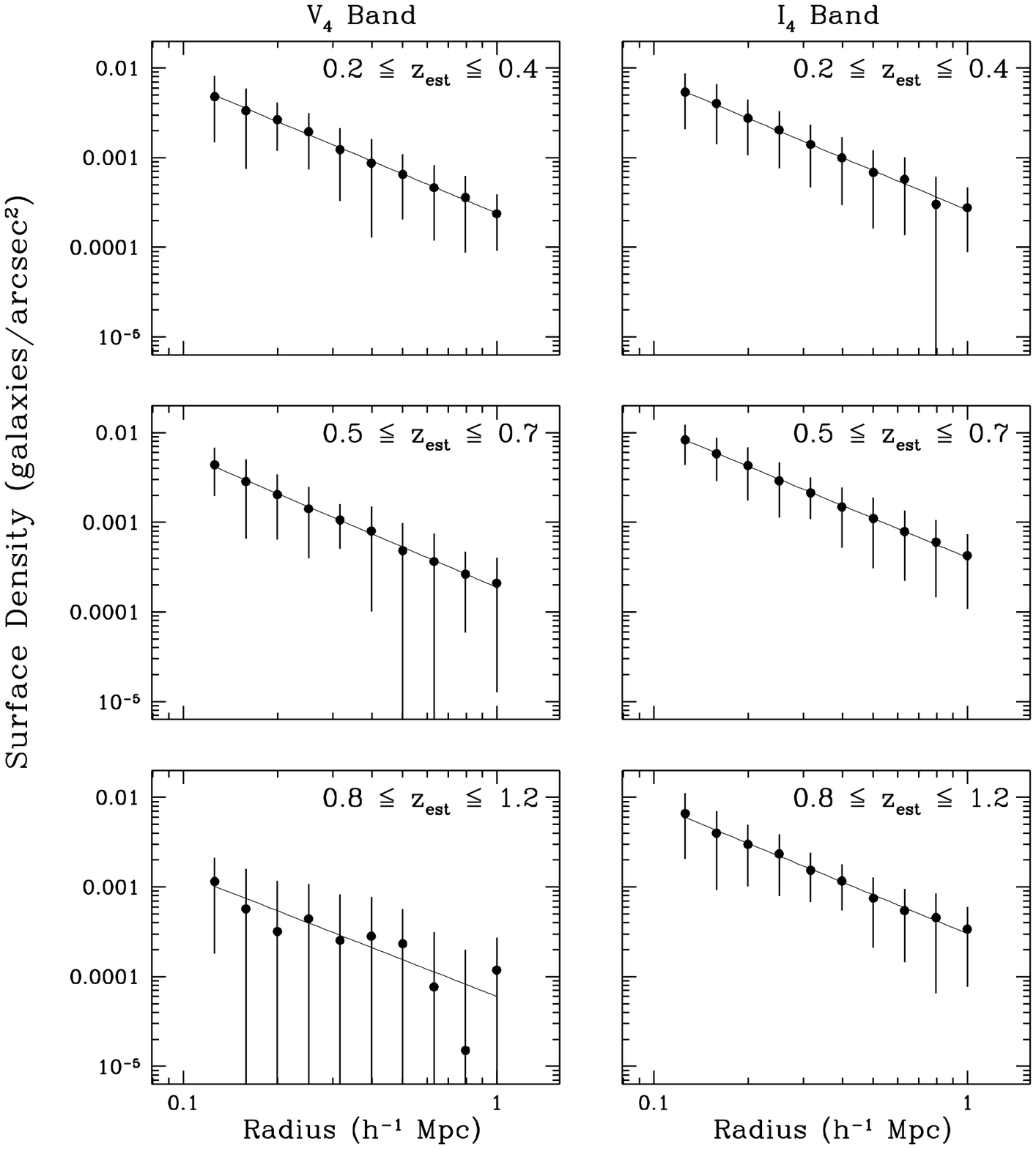}}
\baselineskip 14pt
\caption{Composite surface density profiles
($0.1 \le r \le 1.0~h^{-1}~{\rm Mpc}$) of the simulated clusters.
The clusters have been divided in three redshift bins based upon their
estimated redshifts in each band.  The constant background surface
density has been subtracted from each individual cluster profile
before creating the composite.  The variance in each bin is used to
compute the error in the mean value. The best-fit power-law functions
are shown.}
\end{figure}

% figure 9
\begin{figure}
\centerline{
\epsfysize=7.5in
\epsfbox{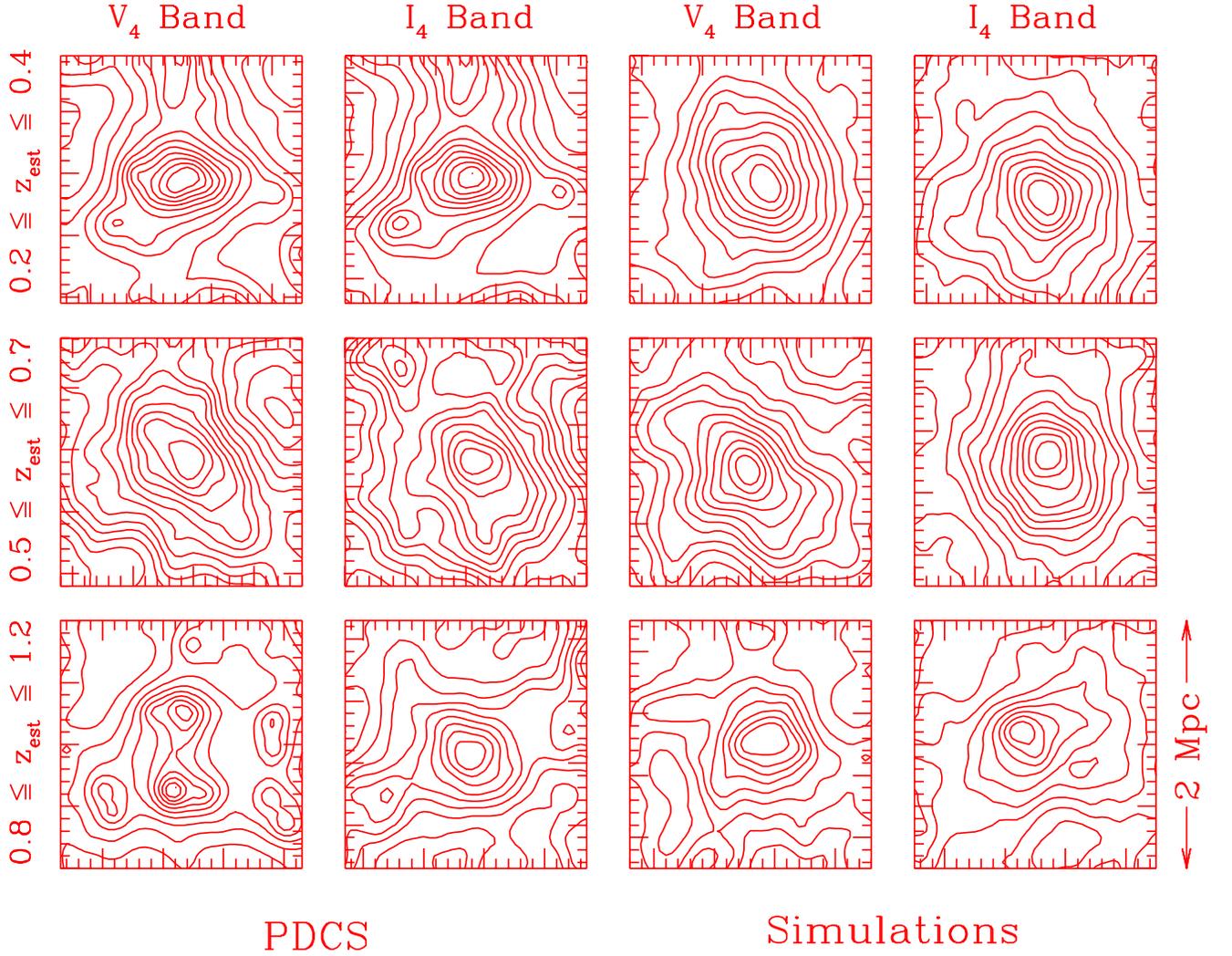}}
\caption{Contour maps of the amplitude of the matched filter signal in
both bands for one cluster in each of the three redshift intervals
(top panel $0.2 \le z_{est} \le 0.4$; middle panel $0.5 \le z_{est}
\le 0.7$; bottom panel $0.8 \le z_{est} \le 1.2$).  The first two
columns show actual PDCS clusters, and the last two columns show
azimuthally symmetric, simulated clusters. The side of each panel
corresponds to $2~h^{-1}~{\rm Mpc}$ at the estimated redshift of the
cluster.}
\end{figure}

% figure 10
\begin{figure}
\centerline{
\epsfysize=7.5in
\epsfbox{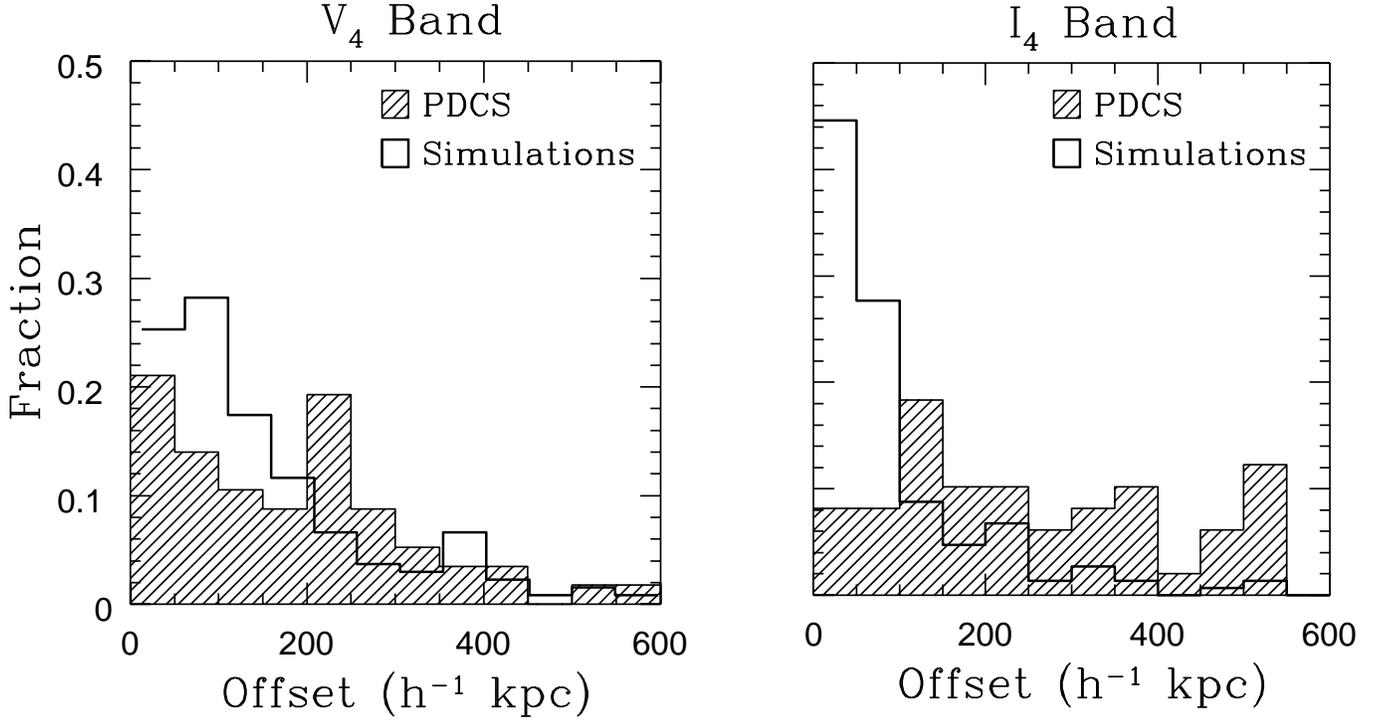}}
\caption{Distribution of offsets (in $h^{-1}~{\rm kpc}$) between the peak
center (the position of the maximum signal of the matched filter) and
the weighted center.  The shaded histograms represent the distribution
of offsets of the PDCS clusters. The solid line histograms represent
the distribution of offsets of the azimuthally symmetric, simulated
clusters. The distributions are shown separately for the \v4 and \i4
bands. The actual data and the simulations are statistically different
(see \S 5).}
\end{figure}

% figure 11
\begin{figure}
\centerline{\epsfbox{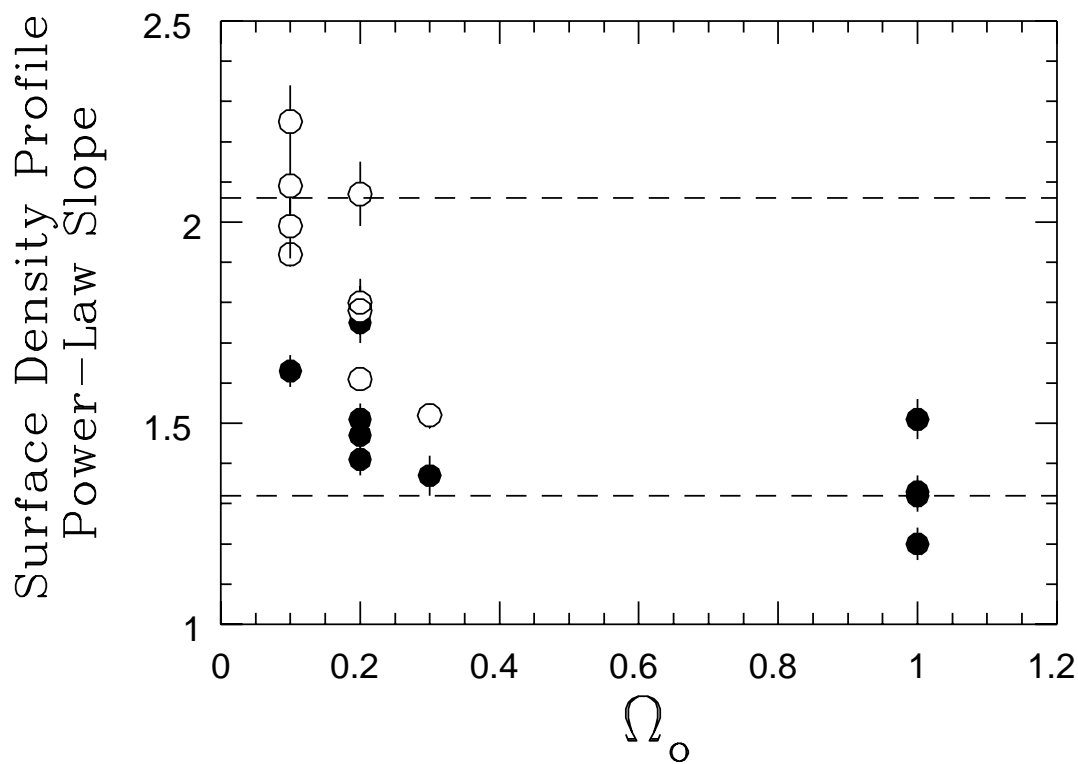}}
\caption{The expected power-law slopes of the cluster {\it mass}
surface density profiles versus $\Omega_{o}$ from various cosmological
simulations (Crone et al. 1994; Jing et al. 1995).  Open and closed
circles indicate open and flat cosmologies, respectively, and
represent simulated clusters at $z = 0$.  Dashed lines indicate the
range of power-law slopes of the composite PDCS cluster {\it galaxy}
profiles over our full redshift interval $0.2 \simless z \simless 1.2$
(Table 5).}
\end{figure}

%
% END 
% 

\end{document}